# A New Paradigm in IBR Modeling for Power Flow and Short Circuit Analysis

Zahid Javid, Firdous Ul Nazir *Senior Member, IEEE*, Wentao Zhu, Diptargha Chakravorty, Ahmed Aboushady *Senior Member, IEEE*, Mohamed Galeela

*Abstract*— The fault characteristics of inverter-based resources (IBRs) are different from conventional synchronous generators. The fault response of IBRs is non-linear due to saturation states and mainly determined by fault ride through (FRT) strategies of the associated voltage source converter (VSC). This results in prohibitively large solution times for power flows considering these short circuit characteristics, especially when the power system states change fast due to uncertainty in IBR generations. To overcome this, a phasor-domain steady state (SS) short circuit (SC) solver for IBR dominated power systems is proposed in this paper, and subsequently the developed IBR models are incorporated with a novel Jacobian-based Power Flow (PF) solver. In this multiphase PF solver, any power system components can be modeled by considering their original non-linear or linear mathematical representations. Moreover, two novel FRT strategies are proposed to fully utilize the converter capacity and to comply with IEEE-2800 2022 std and German grid code. The results are compared with the Electromagnetic Transient (EMT) simulation on the IEEE 34 test network and the 120 kV EPRI benchmark system. The developed IBR sequence domain PF model demonstrates more accurate behavior compared to the classical IBR generator model. The error in calculating the short circuit current with the proposed SC solver is less than 3%, while achieving significant speed improvements of three order of magnitudes.

*Index Term*s—Fault Ride Through, IBR Fault Behavior, Phasor-domain Solver, Power Flow, Short Circuit Solver

## I. Introduction

Inverter-based resources (IBRs) respond differently to faults compared to synchronous generators. To comprehend how IBRs' distinct fault response affects protection system performance, it is crucial to characterize their fault current. This understanding enables more accurate modeling of IBRs in Short Circuit (SC) studies and aids in designing reliable protection schemes for systems dominated by IBRs. The fault response of IBRs depends on their control. Current grid codes and standards require IBRs connected to the bulk power system to withstand various disturbances, including SC faults, and to support the grid. Consequently, diverse Fault Ride-Through (FRT) strategies have been developed. While grid codes and standards outline general performance requirements for FRT behavior, the specifics are mainly left to manufacturers.

This work is supported by Knowledge Transfer Project by Innovate UK (TNEI-13688) between Glasgow Caledonian University, Glasgow and TNEI Services Ltd, United Kingdom (UK).

Z. Javid, is with Department of Electrical and Electronics Engineering, Glasgow Caledonian University, Glasgow, and TNEI Services Ltd, Manchester, UK: (zahid.javid@gcu.ac.uk); F. Ul Nazir, A. Aboushady are with Department of Electrical and Electronics Engineering, Glasgow Caledonian University, Glasgow UK: (FirdousUl.Nazir@gcu.ac.uk; Ahmed.Aboushady@gcu.ac.uk); W. Zhu is with TNEI Services Ltd, Manchester,UK;wentao.zhu@tneigroup.com); M. Galeela is with Faculty of Engineering Cairo University, Egypt and TNEI Services Ltd, Manchester, UK (mohamed.galeela@tneigroup.com) D. Chakravorty is with Siemens Energy Digital Grid, Germany (diptarghachakravorty@gmail.com)

This often leads to proprietary and non-standardized fault current characteristics for IBRs. The IBRs have rapid dynamics and full controllability during disturbances. However, they must limit the current when overloaded [1]. This current limitation alters converter operation and should be considered in SC calculations.

In IBR dominated power systems, accurate SC calculations facilitate the optimal design of power converters for grid-support operations. Numerous studies have examined the fault response of IBRs [2-4]. These efforts are driven by changes in the SC characteristics of power systems due to the increasing integration of IBRs. The industry needs to understand these differences, model IBRs in fault analysis solvers, and design protective relays to maintain the reliability of protection systems in grids with IBRs. Current grid codes (German: VDE-AR-N- 4120, 4110, 4105) and standards (IEEE-2800 and IEEE-1547) mandate that IBRs connected to the bulk power system must ride through various disturbances, including SC faults, while continuing to support the grid. Consequently, various FRT strategies have been developed and integrated into the control systems of IBRs.

The synchronous generators are regarded as the primary source SC current and are typically modeled as a voltage source behind Thevenin impedance for SC calculations [5]. However, this linear equivalent is inadequate for representing IBRs due to the complex controls and saturation operation characteristics of Voltage Source Converter (VSC) [6]. The IEC 60909 standard specifies that IBRs should be modeled as current sources that inject maximum current for SC calculations [7]. In specific fault scenarios, the determination of the current angle, considering the converter control mode, remains unclear. This standard exclusively pertains to Grid Following (GFL) controller. The Grid Forming (GFM) controller can be modeled as controlled voltage sources or as voltage sources with Virtual Impedance Control (VIC); however, other strategies also exist, as discussed in this paper, which reflect a broader range of practical implementations. The GFL controller generally maintains only the current control loop, with current references set according to the specific grid code to function as a voltage-controlled current source [8, 9]. Certain grid codes require the GFL controller to inject only positive sequence positive sequence reactive current, with the remaining capacity used to generate positive sequence active current. However, some protective relays may encounter difficulties if negative sequence current is not injected [10]. The negative sequence voltage during unbalanced faults can cause double-frequency power oscillations. New control strategies for decoupled sequence control (DSC) based controllers have been developed to eliminate these oscillations [11].

According to IEEE-2800 2022 std, the injected negative sequence current must lead the negative sequence voltage by 90° to 100°. This standard also requires the IBR unit to prioritize the reactive component of the current over its active component. For balanced faults this is straightforward as only positive sequence components are present. For unbalanced faults, prioritization needs to account for both positive sequence and negative sequence reactive components. Clause 7.2.2.3.4 of IEEE-2800 2022 std states that: "if the IBR unit's total current limit is reached, either ΔIR1, or ΔIR2, or both may be reduced with a preference of equal reduction in both currents. Additionally, the ΔIR1 injection shall not be reduced below ΔIR2." Here ΔIR1 (positive sequence) and ΔIR2 (negative sequence) represent the active and reactive current contributions of the IBR, respectively. However, this standard does not mention the ratio of injected negative sequence current to positive sequence current. On the other hand, German grid code provides the information of the value of negative sequence current with k-factor but is unclear about the angle. In accordance with the German grid code, the negative sequence control loop is also represented as a controlled current source [12]. In [23], the GFM controller employs a current limiting loop to limit the fault current, functioning as a voltage source with adjustable impedance [13]. However, this model fails to accurately represent the feature of controllable current source. If the current limiting loop is active, the GFM controller can be modeled as a controlled current source, aligning the phasor-domain model with that of the GFL controller [14]. In addition to using the current limiting loop, VIC can effectively limit the fault current in the GFM controller [15]. This method maintains the voltage source characteristic and allows for designing the virtual impedance angle. The positive sequence control loop operates as a voltage source with adjustable virtual impedance [16]. Reference [17] also considered the negative sequence current injection. However, the phasor-domain model is not well-established for this FRT strategy when calculating the steady state (SS) fault current. To the best of the authors' knowledge, there is no established phasor-domain model for adaptive virtual impedance available in the literature.

In the literature, SC analysis for IBR dominated power systems is predominantly conducted through dynamic studies. A quasi SS model has been developed to ascertain the SC current characteristics of VSC, incorporating control dynamics [18]. The Electromagnetic Transient (EMT) simulations can accurately analyze the fault current and voltage during transient and steady states under various FRT control strategies. However, dynamic studies are more expensive and time-consuming than the SS approach. The SS-SC calculations have also been reported in the literature. Authors in [19, 20] recommend modeling the VSC as a current source that injects maximum current, with the current angle estimated based on the pre-fault operating point. References [12, 21, 22] proposed a VSC phasor-domain model where the current injection is iteratively adjusted, considering grid-support control and converter current limitations for SC calculations. However, these are limited to VSCs in GFL control, specifically in PQ mode. The IBRs should be modeled accurately in the power flow (PF) solver which sets the basis for the phasor-domain SC solver. The IBRs cannot be modeled as synchronous generators (voltage source behind Thevenin impedance) in SC solver and have different representation for Grid Following (GFL) and Grid Forming (GFM) modes for positive sequence and negative sequence networks. Also, the IBRs cannot be modeled as PQ generators in the PF solver for decoupled control. A more extensive investigation encompassing diverse converter control modes, such as GFM resource remains necessary for accurate SC calculations in SS solvers. Additionally, protection engineers often focus solely on fault current and voltage during the fault SS. Thus, establishing a SS-SC model for these controllers is crucial. The objective is to develop a phasor domain solver that has comparable accuracy to an EMT solver at a fraction of its computational time.

The main contributions of this paper are listed below:
- Accurate representation of IBRs with sequence domain models PF models into existing modified augmented nodal analysis (MANA) formulation and their integration to SC solver.
- Improved FRT strategies to: 1) fully utilize the SC capacity of the grid-side inverter, ensuring the negative sequence current complies with the German grid code and IEEE-2800 2022 std 2) Improved sequence current distribution and convergence.
- Phasor-domain SS-SC solver featuring developed FRT strategies, IBR models (SS and PF) and multiphase PF solver.

The rest of the paper is organized as follows: Section II develops the sequence domain IBR SC models outlined the proposed FRT strategies. In Section III, the IBR sequence domain PF model and PF formulation are developed. Section IV outlines the proposed SS-SC solver. Test network details and simulation results are discussed in Section V. Finally, conclusions are drawn in Section VI.

## II. IBRs SEQUENCE DOMAIN MODELS AND FRT STRATEGIES

Fig. 1 illustrates the SS sequence domain models of an IBR unit. In the positive sequence, the GFL resource is represented as a voltage-controlled current source. Conversely, in the negative sequence, it can be modeled as either a negative sequence impedance (reactance) or a negative sequence voltage-controlled current source. The negative sequence current or impedance is adjustable based on the chosen Fault Ride Through (FRT) strategy. For example, to adhere to the German grid code, this value can be configured according to the k-factor [23]. The k-factor is used to determine the required amount of reactive power support that an IBR unit must provide relative to its active power output.

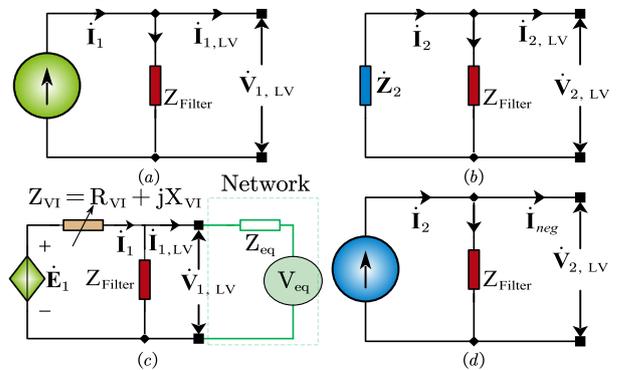

Fig. 1. IBR sequence domain models: (a) positive sequence GFL; (b) negative sequence GFL; (c) positive sequence GFM with VIC; (d) negative sequence GFL and GFM.

GFM is modeled as a voltage source with virtual impedance in the positive sequence. In the positive sequence, the IBR acts as a voltage source to maintain grid stability and support voltage regulation. This helps in providing a stable voltage reference and supporting the grid's active power control. Conversely, in the negative sequence, it is represented as a voltage-controlled current source since in negative sequence, which typically arises from unbalanced conditions, the IBR is represented as a current source.

*A. GFM Control and Current Limiting Strategies*

This section presents a comprehensive review of FRT strategies for GFM available in the literature and introduces a novel FRT approach that ensures compliance with both IEEE-2800 2022 std and German grid code requirements. While the FRT strategies for GFL technology have been extensively documented in existing literature [24, 25], they fall outside the scope of this paper. For GFL implementation in this work, a conventional current saturation method (CSM) is employed. Fig. 2 illustrates the typical control system of a GFM in the *dq* reference frame. This system consists of two layers: the outer control loop and the inner control loop. The primary goal of the outer control loop is to synchronize the GFM resource with the power grid and regulate the terminal voltage magnitude and angle. Common configurations for the outer control loop include droop, virtual synchronous machine (VSM), dispatchable virtual oscillator (dVOC), and phase locked loop (PLL) as shown in Fig. 2. It is important to note that the droop control, dVOC, and VSM are designed to operate during normal and emergencies conditions as they inherently provide GFM functionality by regulating the voltage magnitude, frequency and synchronization under all operating conditions. Unlike PLL, which is generally not used during normal operating conditions but is activated only in emergency scenarios, such as during faults or blackstart operation, for synchronization.

The control system employs a hierarchical signal flow where $E_{ref}$ from the outer control loop establishes the $\dot{V}_{ref}^{d+}$. Meanwhile, the $\dot{V}_{ref}^{q+}$ is kept at zero to maintain proper alignment with the grid voltage reference frame. The voltage control loop processes these voltage references to generate corresponding current setpoints ($\dot{I}_{ref}^{d+}$ and $\dot{I}_{ref}^{q+}$) for the GFM terminals. These current references are then fed into the current control loop which ultimately produces the voltage modulation signal ($V_{PWM}$) at the inverter terminal. During a fault, the GFM control tries to maintain its internal voltage vector, which can push the required $\dot{I}_{ref}^{d+}$ and $\dot{I}_{ref}^{q+}$ above the inverter's maximum current limits. To protect the inverter and ensure it can ride through the fault, additional control loops are integrated into the GFM control system. These additional control loops are activated only during fault to manage the FRT process (see Fig. 3). The current limiting approach plays a vital role in the FRT strategy by keeping the $\dot{I}_{ref}^{d+}$ and $\dot{I}_{ref}^{q+}$ within specified maximum thresholds during fault conditions.

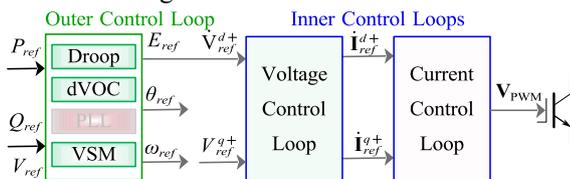

Fig. 2. GFM Control during normal operating conditions

The FRT strategy can incorporate additional goals beyond current limiting, such as supporting grid voltage and frequency stability, while complying with the host grid code's FRT requirements. The way a GFM resource behaves during faults is primarily determined by its FRT strategy.

Current limiting strategies in GFM inverter during fault conditions remain a topic of active research, with no universal agreement on the best approach [26]. While GFL inverters typically rely on current saturation and PLL based synchronization during faults, some GFM schemes adopt similar methods, switching to PLL synchronization when fault conditions exceed the inverter's operational limits [27]. However, GFM inverters also offer unique possibilities for fault management that differ significantly from GFL schemes.

The existing literature identifies two main classifications of FRT strategies for GFM. 1- The converter switches from GFM mode to GFL mode (e.g., CSM), effectively operating as a current source with built-in current limits during fault [28]. For instance, during severe faults, the inverter may prioritize maintaining its internal voltage vector within safe limits while saturating the current [29]. 2- The converter stays in GFM mode during fault conditions (e.g., VIC) using specific current limiting methods while maintaining voltage source behavior [30]. The second strategy maintains the GFM control structure operational during FRT and employs a current limiting technique to prevent overcurrent. Different current limiting strategies are studied in literature such CSM [14, 30], VIC [8, 31], Voltage Limiter Method [32, 33], and hybrid approaches [17, 34]. A high-level overview of these current limiting strategies is given in Fig. 3.

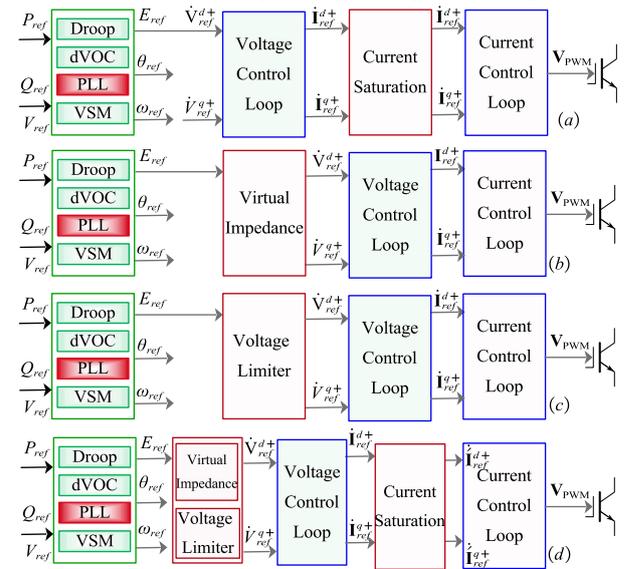

Fig. 3. Current limiting strategies for GFM: (a) Current Saturation Method (CSM); (b) Virtual Impedance Control (VIC); (c) Voltage limiter method; (d) Hybrid method.

The CSM, implemented at the current control loop stage, offers high-speed current limiting due to its higher bandwidth compared to the voltage control loop. This allows it to limit inverter current faster than VIC and VLM, which operate at the voltage control stage. The VIC and VLM struggle with speed due to lower bandwidth which can lead to temporary overcurrent during initial fault cycles. A hybrid method combines these approaches to enhance overall performance. For example, utilizing CSM during transients and VIC post-transients ensures rapid current limiting while maintaining effective voltage regulation during faults. Since the proposed solver is an SS solver, it is limited to observing post-transient impacts. While the proposed strategies can be applied in a hybrid approach for EMT studies, this is beyond the scope of this paper. Instead, the focus here is to establish a phasor-domain steady state model for CSM and VIC with improved features, optimizing the benefits of each technique.

It is important to note that this study assumes the outer loop dynamics are slower than the fault dynamics. Consequently, for steady state SC calculations, the outer loop can be neglected without a significant loss of accuracy, as the short-circuit current magnitude is predominantly governed by the inner loop and hardware limitations [26, 27, 29].

*B. Current Saturation Method (CSM)*

The conventional CSM equations are given below. The detail of these equations can be found in [17, 35]. In this paper all sequence quantities are represented with a dot(.) on top and all phasor quantities are represented with arrow (→) on top. Moreover, zero, positive and negative sequence quantities are represented with subscripts 0, 1 and 2 respectively.

$$|\dot{I}_1| + |\dot{I}_2| = \sqrt{\dot{I}_{1P}^2 + \dot{I}_{1R}^2} + \dot{I}_{2R} \leq \vec{I}_{\lim} \quad (1)$$

$$\dot{I}_{1R,\lim} = \dot{I}_{1R} \frac{\vec{I}_{\lim}}{\dot{I}_{1R} + \dot{I}_{2R}} \quad (2)$$

$$\dot{I}_{2R,\lim} = \dot{I}_{2R} \frac{\vec{I}_{\lim}}{\dot{I}_{1R} + \dot{I}_{2R}} \quad (3)$$

$$\dot{I}_{1P} = \sqrt{(\vec{I}_{\lim} - \dot{I}_{2R})^2 - \dot{I}_{1R}^2} \quad (4)$$

where, $\dot{I}_1$ and $\dot{I}_2$ are positive sequence and negative sequence currents, respectively; $\vec{I}_{lim}$ is the total current limit of the converter; $\dot{I}_{1P}$ and $I_{IR}$ are positive sequence active and reactive current, respectively; $\dot{I}_{2R}$ is the negative sequence reactive current.

Reference [17, 35] introduces a refined CSM with three additional steps to comply with IEEE-2800 2022 std.
1. Setting the total current limit to the maximum phase current.
2. Prioritizing the $\dot{I}_{1P}$ over the $\dot{I}_{1R}$ and $\dot{I}_{2R}$.
3. Ensuring the condition that $\dot{I}_{1R} > \dot{I}_{2R}$ is maintained.

The relationship between three-phase currents sequence currents can be written as follows:

$$|\vec{I}_{abc}| = \sqrt{\left(|\dot{I}_1|^2 + |\dot{I}_2|^2 + 2|\dot{I}_1||\dot{I}_2| \cdot \cos \Delta \vec{\delta}_{abc}\right)} \quad (5)$$

$$\Delta \vec{\delta}_{abc} = \begin{bmatrix} \dot{\delta}_{i1} - \dot{\delta}_{i2} \\ \dot{\delta}_{i1} - \dot{\delta}_{i2} + 2\pi/3 \\ \dot{\delta}_{i1} - \dot{\delta}_{i2} - 2\pi/3 \end{bmatrix} \quad (6)$$

Where $\dot{\delta}_{i1}$ and $\dot{\delta}_{i2}$ are positive sequence and negative sequence current angles at the inverter terminal. To comply with IEEE-2800 2022 std, the negative sequence current at the inverter terminal should lead the negative sequence voltage by 90°, so $\dot{\delta}_{i2}$ can be calculated as:

$$\dot{\delta}_{i2} = \dot{\delta}_{v2} + \pi/2 \quad (7)$$

where $\dot{\delta}_{v2}$ is the measured negative sequence voltage angle at the low voltage (LV) side of the step-up transformer.

According to positive sequence and negative sequence current limiting value distribution method proposed in [17] using the ratio $\eta_{12}$ (positive sequence / negative sequence current), the current limiting values $\dot{I}_{1max}$ and $\dot{I}_{2max}$ can be calculated as follows:

$$\begin{cases} \dot{I}_{2\max} = \dot{I}_{\max} \left[\eta_{12}^2 + 2\eta_{12} \max(\cos(\Delta \vec{\delta}_{abc})) + 1\right]^{-1/2} \\ \dot{I}_{1\max} = \eta_{12} \dot{I}_{2\max} \end{cases} \quad (8)$$

where $\dot{I}_{1max}$ and $\dot{I}_{2max}$ are the maximum values of positive sequence and negative sequence current respectively, $\dot{I}_{max}$ is total maximum current, $\eta_{12}$ positive sequence to the negative sequence current ratio.

However, $\dot{I}_2$ may not reach $\dot{I}_{2max}$ for remote-end faults or high resistance faults. Therefore, the maximum value among three-phase currents will still be lower than $\dot{I}_{max}$. Two improvements are made in existing CMS method: 1) the reactive current is transformed to the *dq*-axis as shown in Fig. 4 (help strictly comply with IEEE-2800 2022); 2) the sequence current distribution is achieved by limiting the $\dot{I}_{2max}$ at $\dot{I}_{max}/2$, which results in better current distribution and convergence.

$$\dot{I}_{2\max} = 0.5 \dot{I}_{\max} \quad (9)$$

Considering the angle difference between $\dot{I}_1$ and $\dot{I}_2$, the $\dot{I}_1$ will always be larger than $\dot{I}_2$ (IEEE-2800 2022 std). Moreover, this distribution will make sure that the converter's full capacity is utilized. Let the maximum value among three-phase currents in (5) equal to $\dot{I}_{max}$ and substitute the actual negative sequence current, the corresponding $\dot{I}_{max}$ can be computed as follows:

$$\dot{I}_{1\max} = -|\dot{I}_2| \max(\cos \Delta \delta_{abc}) \\ + \sqrt{|\dot{I}_2|^2 \cdot (\max(\cos \Delta \delta_{abc}))^2 - |\dot{I}_2|^2 + \dot{I}_{\max}^2} \quad (10)$$

*C. Virtual Impedance Control (VIC)*

This method involves adding a large virtual impedance between the inverter and the grid during an overcurrent event to limit the inverter current. As shown in Fig. 3, the implementation occurs at the voltage control loop stage, where the voltage drops across the virtual impedance is subtracted from the outer control loop's reference voltage $\dot{E}_{ref}$.

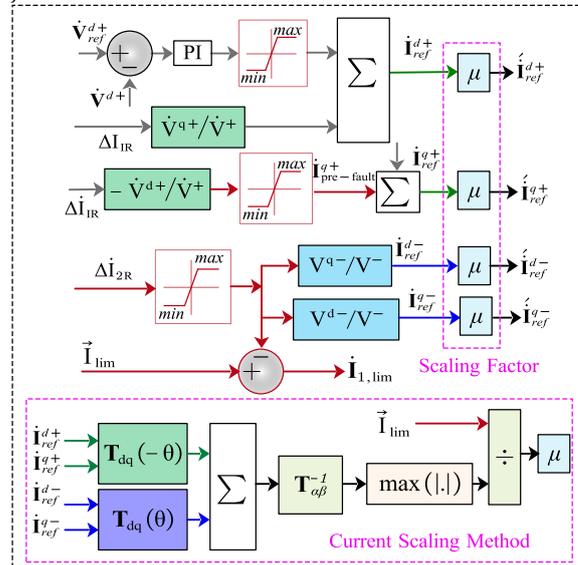

Fig. 4. Improved CSM with current scaling

Consequently, the reference voltages $V_{ref}^{d+}$ and $V_{ref}^{q+}$ are lowered, reducing the effective voltage at the inverter terminal and thus limiting the current.

$$\dot{V}_{ref}^+ = Z_{VI} \dot{I}_1 \rightarrow \begin{cases} \dot{V}_{ref,VI}^{d+} = R_{VI} \dot{I}^{d+} - X_{VI} \dot{I}^{q+} \\ \dot{V}_{ref,VI}^{q+} = R_{VI} \dot{I}^{q+} + X_{VI} \dot{I}^{d+} \end{cases} \quad (11)$$

where: $R_{VI} = R_{VI}^0 + \Delta R_{VI}$; $X_{VI} = X_{VI}^0 + \Delta X_{VI}$, $R_{VI}^0$ and $X_{VI}^0$ are the rated resistance and reactance for load sharing, and they are usually taken as zero during a fault; $\Delta R_{VI}$ and $\Delta X_{VI}$ are their additional components during a fault, and they are adjusted adaptively according to the fault current.

$$\begin{cases} \Delta R_{VI} = \max\left\{\sigma(\sqrt{(\dot{I}_{ref}^{d+})^2 + (\dot{I}_{ref}^{q+})^2} - \vec{I}_{TH}), 0\right\} \\ \Delta X_{VI} = \Delta R_{VI} \cdot \varphi \end{cases}$$

where $\dot{I}_{TH}$ is the current threshold value, and $\varphi$ is the ratio of $X_{VI}$ to $R_{VI}$ (X/R ratio). In addition, $\sigma$ can be expressed as:

$$\sigma = \frac{\vec{v}_{drop}}{\dot{I}_{1\max} \dot{I}_{TH} \sqrt{1+\varphi^2}}, \dot{I}_{TH} = (\dot{I}_{1\max} - \kappa) \qquad (12)$$

where $\dot{v}_{drop}$ is the voltage drop on the virtual impedance, and $\kappa$ is the current threshold gap.

To limit the positive sequence fault current, $\dot{v}_{drop}$ must be adjusted according to the output positive sequence fault current. The X/R ratio of virtual impedance is key to balancing stability, current limitation, and voltage support. A lower X/R ratio can reduce oscillations and overcurrent risk but may weaken voltage support. High virtual impedance keeps inverter current within the limit but may increase instability risk. Mostly the existing VIC methods in literature are designed with a constant impedance to limit inverter current to a fixed threshold, typically the maximum current limit of power electronic switches. While this approach can effectively limit inverter current during symmetrical faults, it falls short for asymmetrical faults. In contrast, an adaptive VIC method can limit current during all types of faults. In the proposed VIC phasor-domain model presented in this paper, $Z_{VI}$ can be calculated based on fault conditions. The contributions regarding VIC are: 1) the development of a sophisticated phasor-domain model; 2) the sequence current distribution ($\dot{I}_{2\max}=I_{\max}/2$); 3) accurate representation of GFM (with VIC) in positive and negative sequence. To adaptively adjust $Z_{VI}$ based on fault conditions, ensuring $|\dot{I}_1|$ is fixed at $\dot{I}_{1\max}$, it is necessary to compute the positive sequence equivalent circuit of the connected network (Fig. 1-c). The $\dot{Z}_{eq}$ and $\dot{V}_{eq}$ can be computed by short-circuiting all the voltage sources and open circuiting all current sources within the network, respectively.

$$\dot{Z}_{eq} = \frac{\dot{V}_{1,LV}}{\dot{I}_{1,LV}}\bigg|_{V_n=0} \; ; \; V_{eq} = \dot{V}_{1,LV}\bigg|_{i_{I,LV}=0} \qquad (13)$$

There are two possibilities depending on the fault: 1- $|\dot{I}_1| < \dot{I}_{1\max}$; 2- $|\dot{I}_1| > \dot{I}_{1\max}$.

In first case the VIC will be inactive and the $\dot{I}_{1,LV}$ can be calculated as follows:

$$\dot{I}_{1,LV} = \dot{I}_1 - \dot{E}_1 / Z_{Filter} \qquad (14)$$

In second case the VIC will be triggered.

$$\dot{I}_1 = \frac{\dot{E}_1 - \dot{Z}_{sum}/\dot{Z}_{eq} \cdot \dot{V}_{eq}}{\dot{Z}_{sum} + \dot{Z}_{VI}} \quad \therefore \dot{Z}_{sum} = \frac{\dot{Z}_{Filter} \cdot \dot{Z}_{eq}}{\dot{Z}_{Filter} + \dot{Z}_{eq}} \qquad (15)$$

Let $|\dot{I}_1|$ equal to $I_{1\max}$, the $R_{VI}$ can be computed as follows:

$$\begin{cases} \dot{R}_{VI} = \frac{-(\dot{R}_{sum} + \varphi \dot{X}_{sum}) + \sqrt{(\dot{R}_{sum} + \varphi \dot{X}_{sum})^2 - (1+\varphi^2)\zeta^2}}{(1+\varphi^2)} \\ \zeta^2 = R_{sum}^2 + X_{sum}^2 - \frac{\left|\dot{E}_1 - \dot{Z}_{sum}/\dot{Z}_{eq} \cdot \dot{V}_{eq}\right|^2}{\dot{I}_{1\max}^2} \end{cases} \qquad (16)$$

After computing the $R_{VI}$, the $X_{VI}$ can be computed with a suitable ratio factor. The $\dot{I}_1$ can be calculated by (15) and then $\dot{I}_{1,LV}$ can be calculated as follows:

$$\dot{I}_{1,LV} = \dot{I}_1 - \dot{V}_{1,LV} / \dot{Z}_{Filter} \qquad (17)$$

Since negative sequence current references depend solely on the negative sequence voltage at the terminal (German grid code), there's no need to detect the internal structure of the connected network. The $I_{ref}^{d-}$ and $I_{ref}^{q-}$ can be calculated as given in (18) to make $\dot{I}_2$ lead $\dot{V}_2$ by 90° (IEEE-2800 2022 std) while the negative sequence current limit is satisfied.

$$\begin{cases} \dot{I}_{ref}^{d-} = \frac{k\dot{V}^{q-}}{\rho} \\ \dot{I}_{ref}^{q-} = \frac{-k\dot{V}^{d-}}{\rho} \end{cases}, \therefore \rho = \max\left(1, \frac{k\sqrt{(\dot{V}^{q-})^2 + (\dot{V}^{q-})^2}}{\dot{I}_{2\max}}\right) \qquad (18)$$

where, $k$ is the k-factor (German Grid Code).

$$\dot{I}_2 = \sqrt{(\dot{I}_{ref}^{d-})^2 + (\dot{I}_{ref}^{q-})^2} \exp(j \cdot (\angle\dot{E} - \arctan \dot{I}_{ref}^{q-}/\dot{I}_{ref}^{d-})) \qquad (19)$$

And then, $\dot{I}_{2,LV}$ can be computed as:

$$\dot{I}_{2,LV} = \dot{I}_2 - \dot{V}_{2,LV} / \dot{Z}_{Filter} \qquad (20)$$

The reactive power Q can be calculated as follows:

$$Q = imag\left(\dot{V}_{1,LV}(\dot{I}_{1,LV})^* + \dot{I}_{2,LV}(\dot{V}_{2,LV})^*\right) \qquad (21)$$

Finally, $|\dot{E}_1|$ can be updated by the droop control.

$$|\dot{E}_1| = \dot{V}_{ref}^{d+} = \vec{V}_{ref} + k_v(Q_{ref} - Q) \qquad (22)$$

where, $k_v$ is a droop constant.

### III. PF FORMULATION AND JACOBIAN FORMATION

For PF solution modified augmented nodal analysis (MANA) approach is used. This multiphase PF formulation allows to analyze power system with multiple phases, accommodating unbalanced loads and configurations. Its advantage lies in its ability to accurately model and simulate real-world conditions, leading to more precise analysis [36]. Moreover, MANA formulation allows juxtaposition and modelling of arbitrary components with arbitrary constraints and unknowns directly into the Jacobian matrix.

The primary distinction between classical nodal analysis and MANA is that MANA facilitates the direct incorporation of components, which are challenging or impossible to represent with their impedance model in the classical $Y_{bus}$ matrix, into the Jacobian matrix. In MANA, current is also an independent variable, so electronically coupled generators (ECG) also known as IBRs and electrically coupled load (ECL) can be modelled accurately and independently into Jacobian easily. Moreover, MANA exhibits better convergence than classical nodal analysis due to its improved condition number, even for ill-conditioned networks [37, 38]. It is important to highlight that we employ DSC for converters, which inherently decouple the positive and negative sequence components [12]. The key assumption made here is that the dynamics of the DSC are fast enough to ensure effective decoupling of the sequences, even during transient conditions. This assumption is widely adopted in literature, where DSC has been shown to perform well in maintaining independent control of positive and negative sequences under balanced and unbalanced conditions [39]. As a result, the coupling between positive and negative sequences introduced by the converter dynamics is not explicitly accounted for in the PF formulation.

### A. MANA PF Formulation

The basic concept of the MANA based PF formulation is presented in (23), where $Y_n$ is the classical nodal admittance matrix, $A_r$, $A_c$, $A_d$ are the augmented row, column and diagonal elements of the system matrix respectively. These elements are used to model components which are not modelled or cannot be modelled in $Y_n$. These components are known as non-consecutive network elements which are hard to model with their terminal voltage alone. The detailed version of (23) for SS MANA is given in (24) and PF MANA is given in (25).

$$\begin{bmatrix} Y_n & A_c \\ A_r & A_d \end{bmatrix} \begin{bmatrix} \vec{V}_n \\ \vec{I}_x \end{bmatrix} = \begin{bmatrix} \vec{I}_n \\ \vec{V}_x \end{bmatrix} \quad (23)$$

In SS MANA, the loads are converted to constant impedance loads with their nominal voltages and generators modeled as ideal voltage sources. It is worth noting that for PF solution, the full Newton Raphson (NR) method is used to solve (25).

$$\begin{bmatrix} Y_n & V_r^T & D_r^T & S_r^T \\ V_r & 0 & 0 & 0 \\ D_r & 0 & 0 & 0 \\ S_r & 0 & 0 & S_d \end{bmatrix}^{(i)} \begin{bmatrix} \Delta \vec{V}_n \\ \Delta \vec{I}_v \\ \Delta \vec{I}_d \\ \Delta \vec{I}_s \end{bmatrix}^{(i)} = - \begin{bmatrix} f_n \\ f_v \\ f_d \\ f_s \end{bmatrix}^{(i)} \quad (24)$$

$$\begin{bmatrix} Y_n & V_r^T & D_r^T & S_r^T & A_{IL} & A_{IG} & 0 & A_{VR} & A_{IBR} \\ V_r & 0 & 0 & 0 & 0 & 0 & 0 & 0 & 0 \\ D_r & 0 & 0 & 0 & 0 & 0 & 0 & 0 & 0 \\ S_r & 0 & 0 & S_d & 0 & 0 & 0 & 0 & 0 \\ C_L & 0 & 0 & 0 & D_L & 0 & 0 & 0 & 0 \\ C_G & 0 & 0 & 0 & 0 & D_G & 0 & 0 & 0 \\ Y_G & 0 & 0 & 0 & 0 & B_G & Y_{GE} & 0 & 0 \\ C_{RV} & 0 & C_{RI} & 0 & 0 & 0 & 0 & C_{RG} & 0 \\ C_{IBR} & 0 & 0 & 0 & 0 & 0 & 0 & 0 & D_{IBR} \end{bmatrix}^{(i)} \cdots$$

$$\begin{bmatrix} \Delta \vec{V}_n \\ \Delta \vec{I}_v \\ \Delta \vec{I}_d \\ \Delta \vec{I}_s \\ \Delta \vec{I}_L \\ \Delta \vec{I}_G \\ \Delta \vec{E} \\ \Delta \vec{g} \\ \Delta \vec{I}_{IBR} \end{bmatrix}^{(i)} = - \begin{bmatrix} f_n \\ f_v \\ f_d \\ f_s \\ f_L \\ f_G \\ f_E \\ f_R \\ f_{IBR} \end{bmatrix}^{(i)} \quad (25)$$

where, $V_r$: is the ideal voltage sources, $D_r$: Transformers, $S_r$: Switches (open position), $S_d$: Switches (Close position), $C_L$ and $D_L$: Load partial derivatives $C_G$ and $D_G$: Generator partial derivatives $Y_G$ and $B_G$: Generator internal EMF partial derivatives, $C_{RV}$, and $C_{RG}$: Voltage regulator partial derivatives, $C_{IBR}$ and $D_{IBR}$ are the partial derivative entries of IBR constraint $f_{IBR}$ with respect to voltage and current real and imaginary parts respectively. $A_{IL}$, $A_{IG}$, $A_{VR}$, and $A_{IBR}$ are adjacency matrices for load, generators, voltage regulators and IBRs respectively. $f_n$, $f_v$, $f_d$, $f_s$, $f_L$, $f_G$, $f_E$, $f_R$, and $f_{IBR}$, are constraints for KVL, ideal voltage sources, transformers, switches constraint, loads, generators, generator internal EMF, voltage regulators and IBRs respectively.

The constraints related to IBRs are explained only in this paper due to space limitations. The control systems of IBRs manage the power delivered to the network at the Point of Interconnection (POI) by utilizing the *dq* components of the currents. Notably, IBRs exhibit distinct characteristics in positive sequence and NS scenarios. For instance, in GFM control during positive sequence, they can be represented either as a variable voltage source or as a voltage source with virtual impedance. Conversely, in the negative sequence, it still functions as voltage-controlled current sources. Consequently, the constraints of IBRs are modeled in the sequence domain to account for these variations, particularly in relation to SC contributions. The phase IBR is an unknown quantity in MANA. By including IBR current in the vector of state variables (*x*), the simplified system of equations is presented as shown in equation (25). The sequence components for both voltage and current shall be defined as follows:

$$\vec{I}_{012} = \vec{A}^{-1} \vec{I}_{IBR} = [\dot{I}_0 \; \dot{I}_1 \; \dot{I}_2]^T \quad (26)$$

$$\vec{V}_{012} = \vec{A}^{-1} \vec{V}_{POI} = [\dot{V}_0 \; \dot{V}_1 \; \dot{V}_2]^T \quad (27)$$

where A is Fortescue transformation [40] matrix. This transformation is used to convert phasor quantities into symmetrical sequence components and vice versa.

### B. Positive Sequence Constraint

The IBRs regulate the positive sequence current to adjust the injected power. In PQ mode, a GFL resource is constrained by the real and reactive powers of the positive sequence. Conversely, a GFM is constrained by the positive sequence component of the injected real power and the magnitude of the positive sequence voltage. Therefore, three constraint equations related to positive sequence are deduced.

The total injected apparent power of the IBR is given below.

$$\vec{S}_{inj} = -3(\dot{I}_0^* \dot{V}_0 + \dot{I}_1^* \dot{V}_1 + \dot{I}_2^* \dot{V}_2) \quad (28)$$

If $P_r$ is the positive sequence injected power, then the positive sequence active power constraint can be formed as follows:

$$f_{IBR1\_p} = \dot{P}_r + \text{Re}(\dot{S}_1) \quad (29)$$

Similarly, positive sequence reactive power constraint can be formed as follows:

$$f_{IBR1\_q} = \dot{Q}_r + \text{Im}(\dot{S}_1) \quad (30)$$

The above two constraints contribute twenty-four Jacobian entries with twelve entries each in $C_{IBR}$ and $D_{IBR}$ for one IBR unit.

If $\dot{V}_r$ is the desired positive sequence voltage ($\dot{V}_r > 0$), then the constraint for the magnitude of the positive sequence can be formed as follows:

$$f_{IBR1\_v} = \dot{V}_r - |\dot{V}_1| \quad (31)$$

This constraint contributes six entries into $C_{IBR}$, and it does not contribute any entry into $D_{IBR}$ because (31) has no current component.

### C. Negative Sequence Constraint

The inner controller produces controlled converter voltages using DC current references in the *dq* frame, along with voltage and current measurements at the POI terminals.

These measurements are filtered and transformed into the *dq* domain. Due to the phase shift introduced by the filter, the negative sequence component in the converter voltages differs from the negative sequence voltage at the POI. Consequently, negative sequence current circulates in the converter. The amount of this current is minimal compared to the positive sequence current, but it is present in unbalanced networks and should be considered when modeling the IBR in power flow. Additionally, this formulation can be used to design the FRT strategy for injecting negative sequence current to comply with various grid codes. The linear relationship between the negative sequence voltage at the POI and the NS current can be expressed as follows:

$$\dot{K}_{neg}\dot{V}_{neg} = \dot{I}_{neg} \quad (32)$$

where, $\dot{K}_{neg}$ is a complex quantity which is the order of admittance in negative sequence. Its value can also be adjusted to comply with the German grid code.

The constraint equation is divided into real and imaginary parts as follows:

$$f_{IBR2\_r} = \mathrm{Re}(\dot{K}_{neg}\dot{V}_{neg} - \dot{I}_{neg}) = 0 \quad (33)$$

$$f_{IBR2\_i} = \mathrm{Im}(\dot{K}_{neg}\dot{V}_{neg} - \dot{I}_{neg}) = 0 \quad (34)$$

It forms twenty-four Jacobian entries with twelve entries each in $C_{IBR}$ and $D_{IBR}$ for one IBR unit.

### D. Zero Sequence Constraint

Due to the design of converters, the IBR does not contribute to zero sequence current. However, the transformers connecting the IBR to the grid may allow zero sequence current to circulate through their windings. Therefore, the zero sequence current equation must account for this circulating current as well. The constraint equation is divided into real and imaginary parts as follows:

$$f_{IBR0\_r} = \mathrm{Re}(\dot{K}_{zero}\dot{V}_{zero} - \dot{I}_{zero}) = 0 \quad (35)$$

$$f_{IBR0\_i} = \mathrm{Im}(\dot{K}_{zero}\dot{V}_{zero} - \dot{I}_{zero}) = 0 \quad (36)$$

where, $\dot{K}_{zero}$ is zero sequence admittance of the transformer.

It forms eighteen Jacobian entries with nine entries in $C_{IBR}$ and nine entries $D_{IBR}$ for one IBR unit.

### IV. STEADY STATE SHORT CIRCUIT SOLVER

There are three main parts of the proposed circuits-SC solver. 1- PF solver (Non-linear Network); 2- SS PF solver (Linearized Network); 3-IBR SS solver. Once the input data is loaded and processed, the system of equations for the MANA-based PF solver is established. The pre-fault conditions are derived from this PF solution. These conditions are used to linearize the network by converting generators into voltage sources behind impedance, loads into RLC equivalents, and GFL resource into current source and GFM into voltage source with virtual impedance. The SS solution of the linearized network, without faults, matches exactly with the initial PF solution. A sparse node matrix *A* is produced to represent the connected system topology. The switches are used to set different fault types and fault locations. The status of the switch is represented by 0 (open) and 1 (close), respectively. A fault is placed in the network and the data is fed to the IBR SS-SC solver.

All the voltage sources in vector *b* is set to 0, a constant unit $\dot{I}_{1,LV}$ is injected. The $\dot{I}_{2,LV}$ is computed by (20) and $\dot{Z}_{eq}$, by (13). To compute $\dot{V}_{eq}$, $\dot{I}_{1,LV}$ is set to 0, and $\dot{I}_{2,LV}$ is computed with (20). After computing the $\dot{Z}_{eq}$ and $\dot{V}_{eq}$, $\dot{I}_1$ is calculated with (14). The calculation of $\dot{I}_{1,LV}$ depends on the value of the $\dot{I}_1$. If its value is less than $\dot{I}_{1max}$ then VIC will be inactive and the $\dot{I}_{1,LV}$ will be calculated with (17). If its value is greater than $\dot{I}_{1max}$ then VIC will be activated to limit the $\dot{I}_1$ at $\dot{I}_{1max}$ with virtual impedance computed by (16) and then $\dot{I}_{1,LV}$ will be updated with (17). The $\dot{I}_2$ will be calculated by (19), and the value of $\dot{I}_{2,LV}$ depends on the $\dot{I}_2$. If its value is less than $I_{2max}$ then $\dot{I}_{2,LV}$ will be calculated by (20); otherwise, $\dot{I}_2$ will be equal to $\dot{I}_{2max}$ ($\dot{I}_{2max}=0.5\dot{I}_{max}$) and $\dot{I}_{2,LV}$ will be calculated by (20). Finally, $|\dot{E}_1|$ will be updated by (22).

Sequence currents are converted to phasor values using Fortescue transformation and updated into the vector *b*, node voltages are found by solving *Ax = b*. For any fault impedance, the nodal admittance matrix is adjusted without expanding it. Due to the non-linear response of IBRs, equations are solved iteratively, updating IBR currents each time. The process checks voltage changes on IBR terminals and continues until converged or reached the iteration limit. The flow chart of the whole process is shown in Fig. 5.

### V. TEST NETWORK AND SIMULATION RESULTS

To validate the IBR PF model developed in Section III, IEEE 34 bus test network was simulated. This feeder is an actual feeder located in Arizona, USA, with a nominal voltage of 24.9 kV. It has two in-line regulators, unbalanced loading, and shunt capacitors. The original network was modified by adding a 200 kW IBR at bus "824" as shown in Fig. 6. The rest of the network parameters remain the same, the detail of the network can be found in [41]. The model parameters ($\dot{K}_{zero}$ & $\dot{K}_{neg}$) for the proposed IBR-PF model are obtained from Electromagnetic Transient Program (EMTP) time-domain simulation.

It is important to note that in this study we used an aggregated model to represent the Wind Park (WP). This approach simplifies the representation of the WP by combining the behavior of individual turbines into a single equivalent model. The aggregated model retains the dynamic and control characteristics of the entire WP, including the implementation of FRT strategies. Although the collector grid and WTGs inside the WP are represented with their aggregated models, the overall reactive power control structure of the WP is preserved by taking the central WP controller into account.

Fig. 7 compares the voltage profile of the IEEE 34 bus test network of proposed and classical generator IBR-PF (voltage source behind impedance) model with EMTP simulations as reference. The proposed IBR-PF model demonstrates superior accuracy, with a maximum percentage error of 0.0453% in voltage magnitude and an average difference of 0.017%. In contrast, the classical IBR model exhibits a maximum error of 0.176% and an average error of 0.0457%. The generator sequence currents for these models are given in TABLE I. The percentage error for current magnitudes of proposed method for zero, positive and negative sequence is 0.034%, 0.027%, and 0.001%, respectively. In contrast, these errors for classical IBR generator models are 17.75%, 1.81%, and 31.32%, respectively.

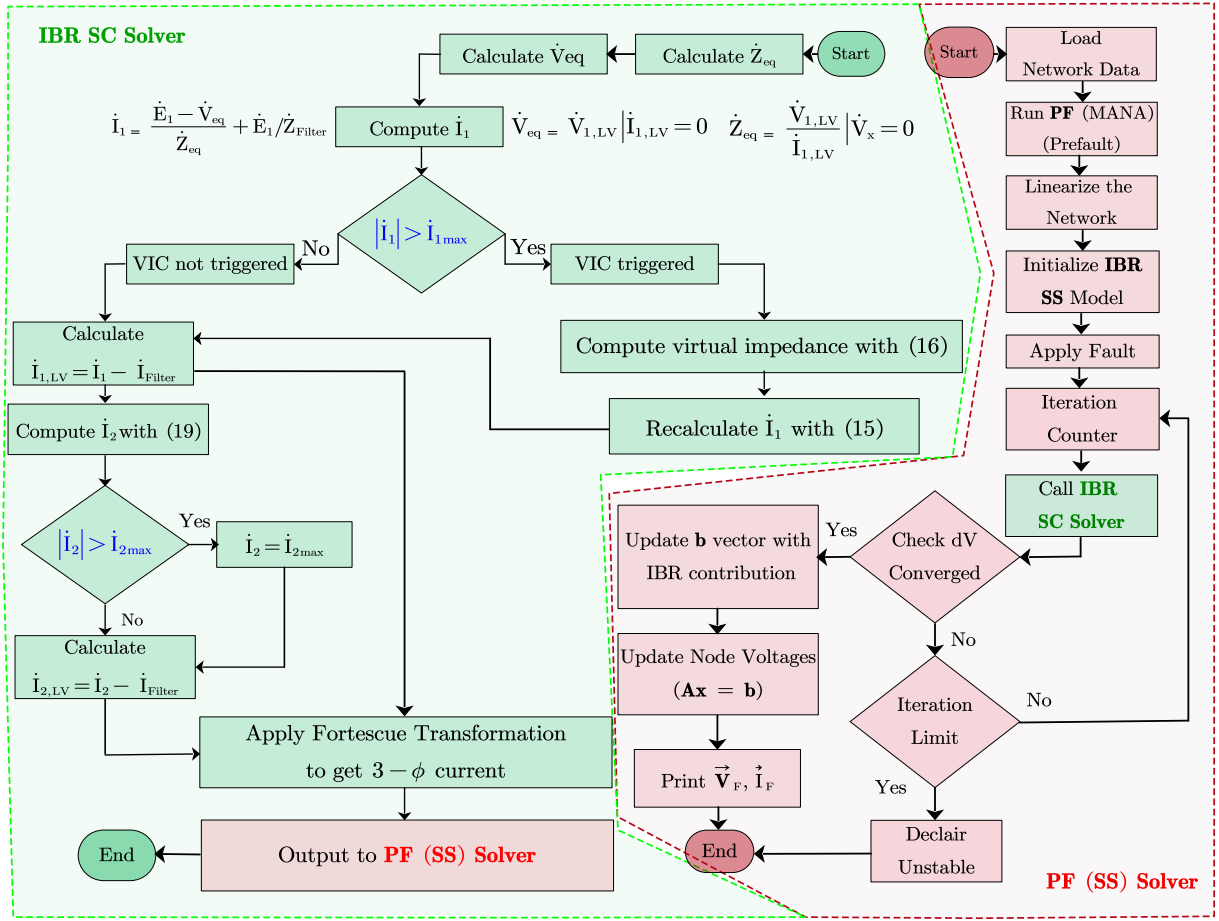

Fig. 5. Proposed SS-SC solver

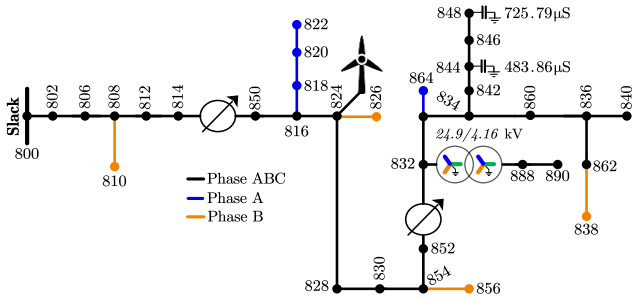

Fig. 6 IEEE 34 bus test network

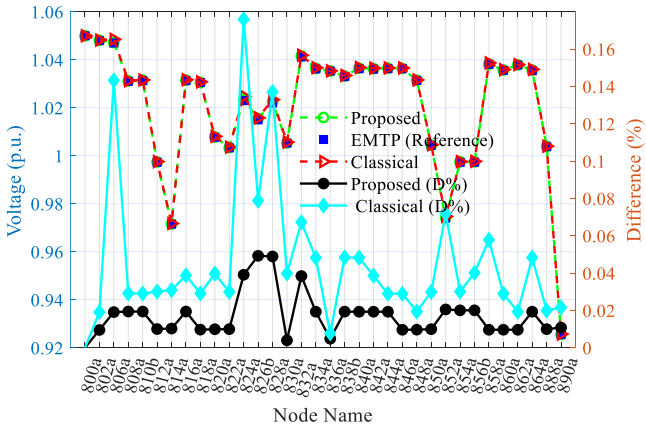

Fig. 7. Accuracy validation of the proposed PF model (D=Difference)

TABLE I. COMPARISON OF GENERATOR SEQUENCE CURRENTS

| | EMTP | | Proposed | | Classical | |
|---|---|---|---|---|---|---|
| Value | Mag. (A) | Angle (°) | Mag. (A) | Angle (°) | Mag. (A) | Angle (°) |
| $\dot{I}_0$ | 1.3415 | 125.56 | 1.3370 | 125.46 | 1.1033 | 126.14 |
| $\dot{I}_1$ | 11.3091 | 178.39 | 11.3122 | 178.41 | 11.5137 | 178.46 |
| $\dot{I}_2$ | 1.5109 | 23.04 | 1.5109 | 22.66 | 1.0377 | 19.71 |

These results confirm that classical generator models cannot accurately represent IBR generators. Since PF solution forms the foundation for SC studies particularly in the context of unbalanced faults where sequence currents are crucial, an accurate IBR-PF model with multiphase PF formulation can significantly enhance the precision of SC solver. To test the performance of the proposed SC solver and FRT strategies, the reduced EPRI 120 kV benchmark, as shown in Fig. 8, is simulated. The test network data can be found in [42]. A line to ground fault (AG) is placed at bus 3. The proposed CSM is compared with the existing CSM method [17] for compliance with IEEE-2800 2022 std. TABLE II shows the phase and sequence currents at IBR terminals.

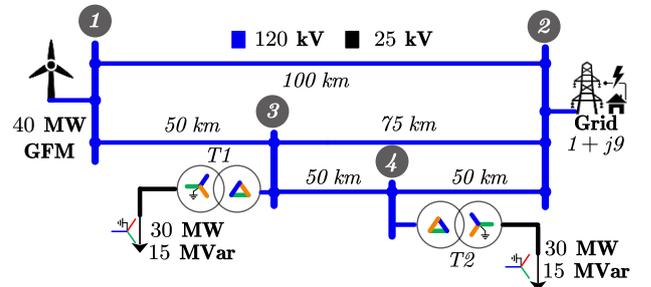

Fig. 8. 120 kV EPRI benchmark (Reduced Circuit)

As illustrated in TABLE II, the proposed CSM allows phase current to reach the current limiting value of (1.1 p.u.). Conversely, in the existing CSM method, the maximum value among the three-phase currents remains below 1.1 per unit, indicating that the converter capacity is not fully utilized. The negative sequence current leads negative sequence voltage by 90°. The results indicate that improved

CSM strategies can fully utilize the converter capacity and strictly comply with IEEE-2800 2022 std. As mentioned earlier, the CSM method loses its voltage source feature during a fault and is only advantageous during transient states when used in hybrid mode with an external current limiting method in outer loop. Therefore, further analysis will focus solely on the developed VIC-based phasor-domain model for GFM, while for GFL resource the proposed CSM is used.

TABLE II. COMPARISON OF CSM FRT STRATEGIES

| Value | EMTP | CSM (Existing) | CSM (Proposed) |
|---|---|---|---|
| $\vec{I}_a$ | 1.100 | 0.991 | 1.100 |
| $\vec{I}_b$ | 0.616 | 0.601 | 0.615 |
| $\vec{I}_c$ | 1.022 | 0.802 | 1.020 |
| $I_{1,LV}$ | 0.959 | 0.780 | 0.957 |
| $I_{2,LV}$ | 0.349(94.20°) | 0.334(90.15°) | 0.348(94.20°) |
| $\Delta\dot{V}_{2,LV}$ | 0.175(4.15°) | 0.157(9.21°) | 0.179(4.21°) |

To test the compliance of the proposed VIC phasor-domain model with the German grid code the $\vec{K}_{neg}$ is set to 2. The results are shown in TABLE III for different faults on Bus 1. The fault BC represents a phase-to-phase fault between phase B and phase C, BCG represents a phase-to-phase-to-ground fault involving Phase B, Phase C, and ground, and ABCG represents a three-phase-to-ground fault involving Phases A, B, C, and ground, respectively. To further validate the robustness and performance boundaries of the proposed phasor-domain SC solver and FRT methodologies, comprehensive simulations were conducted on the standardized EPRI 120kV benchmark system depicted in Fig. 9. The test network data can be found in [42]. The test network incorporates one GFM and two GFL resources operating under distinct control architectures. For GFM, the proposed VIC-FRT strategy was implemented, while GFL utilized the proposed CSM.

TABLE III. ACCURACY VALIDATION OF VIC PHASOR-DOMAIN MODEL

| Type | Value | EMTP | VIC | Error (%) |
|---|---|---|---|---|
| BC | $\vec{I}_a$ | 1.100 | 1.100 | 0.00% |
|  | $\vec{I}_b$ | 0.568 | 0.569 | 0.17% |
|  | $\vec{I}_c$ | 0.576 | 0.574 | 0.35% |
|  | $I_1$ | 0.693 | 0.691 | 0.29% |
|  | $I_2$ | 0.509 (92.2°) | 0.508 (92.1°) | 0.20% |
|  | $\Delta\dot{V}_{1,LV}$ | 0.303 | 0.304 | 0.33% |
|  | $\Delta\dot{V}_{2,LV}$ | 0.251 (2.25°) | 0.252 (2.20°) | 0.40% |
| BCG | $\vec{I}_a$ | 1.100 | 1.100 | 0.00% |
|  | $\vec{I}_b$ | 0.666 | 0.663 | 0.45% |
|  | $\vec{I}_c$ | 0.661 | 0.663 | 0.30% |
|  | $I_1$ | 0.764 | 0.762 | 0.26% |
|  | $I_2$ | 0.435 (93.10°) | 0.438 (93.13°) | 0.68% |
|  | $\Delta\dot{V}_1$ | 0.323 | 0.322 | 0.31% |
|  | $\Delta\dot{V}_2$ | 0.218 (3.10°) | 0.217 (3.12°) | 0.46% |
| ABCG | $\vec{I}_a$ | 1.100 | 1.100 | 0% |
|  | $\vec{I}_b$ | 1.100 | 1.100 | 0% |
|  | $\vec{I}_c$ | 1.100 | 1.100 | 0% |
|  | $I_1$ | 1.100 | 1.100 | 0% |
|  | $I_2$ | 0.000 | 0.000 | - |
|  | $\Delta\dot{V}_1$ | 0.355 | 0.355 | 0% |
|  | $\Delta\dot{V}_2$ | 0.000 | 0.000 | - |

The results yielded from a three phase (ABC) fault simulated at bus 4 are presented in TABLE IV. The close agreement with EMTP demonstrates the effectiveness of the proposed framework in analyzing networks containing multiple IBRs with diverse control schemes.

TABLE IV. ACCURACY VALIDATION FOR MULTI IBR SYSTEM

| WP | Value | EMTP | Proposed | Error (%) |
|---|---|---|---|---|
| WP1 (GFL) | $\vec{I}_a$ | 1.013 | 1.006 | 0.69% |
|  | $\vec{I}_b$ | 1.100 | 1.091 | 0.82% |
|  | $\vec{I}_c$ | 0.917 | 0.911 | 0.65% |
|  | $I_{1,LV}$ | 0.999 | 0.993 | 0.60% |
|  | $I_{2,LV}$ | 0.231 (93.785°) | 0.228 (93.320°) | 1.29% |
|  | $\Delta\dot{V}_{1,LV}$ | 0.342 | 0.337 | 1.46% |
|  | $\Delta\dot{V}_{2,LV}$ | 0.115 (3.819°) | 0.114 (3.578°) | 0.87% |
| WP2 (GFM) | $\vec{I}_a$ | 0.826 | 0.832 | 0.73% |
|  | $\vec{I}_b$ | 1.100 | 1.100 | 0.00% |
|  | $\vec{I}_c$ | 0.843 | 0.848 | 0.59% |
|  | $I_{1,LV}$ | 0.906 | 0.905 | 0.11% |
|  | $I_{2,LV}$ | 0.298 (104.97°) | 0.294 (104.84°) | 1.34% |
|  | $\Delta\dot{V}_{1,LV}$ | 0.103 | 0.102 | 0.97% |
|  | $\Delta\dot{V}_{2,LV}$ | 0.148 (14.998°) | 0.147 (14.837°) | 0.66% |
| WP3 (GFL) | $\vec{I}_a$ | 1.001 | 1.005 | 0.40% |
|  | $\vec{I}_b$ | 1.087 | 1.100 | 1.19% |
|  | $\vec{I}_c$ | 0.969 | 0.954 | 1.55% |
|  | $I_{1,LV}$ | 0.988 | 0.983 | 0.51% |
|  | $I_{2,LV}$ | 0.069 (93.785°) | 0.070 (93.320°) | 1.45% |
|  | $\Delta\dot{V}_{1,LV}$ | 0.041 | 0.040 | 2.44% |
|  | $\Delta\dot{V}_{2,LV}$ | 0.034 (3.819°) | 0.035 (3.578°) | 2.94% |

Different fault scenarios were simulated with standardized EPRI 120kV benchmark system depicted in Fig. 9, for simulation time comparison. The simulations were performed using Lenovo X1 Carbon, ThinkPad, Core i7- CPU @ 1.80 GHz & 2.30 GHz, with 16 GB RAM. TABLE V illustrates the performance enhancements achieved by the proposed phasor-domain SS-SC solver in comparison to EMTP simulations. In all cases the speed gain is more than 1000 times as compared to EMTP simulations with maximum error of 2.96%.

TABLE V. SPEED GAINS

| Case | Fault | Bus | Time (Seconds) EMTP | Time (Seconds) Proposed | Speed Gain (times) | $I_{Fault}$ Error (%) |
|---|---|---|---|---|---|---|
| 1 | ABCG | 8 | 60.234 | 0.051 | 1181 | 0.51 |
| 2 | BC | 4 | 60.681 | 0.049 | 1238 | 1.78 |
| 3 | BCG | 4 | 66.586 | 0.066 | 1008 | 1.18 |
| 4 | AB | 2 | 64.891 | 0.064 | 1013 | 1.04 |
| 5 | ABC | 2 | 61.274 | 0.059 | 1038 | 2.96 |
| 6 | ACG | 10 | 62.077 | 0.062 | 1001 | 1.17 |

Fig. 10, Fig. 11 and Fig. 12 show the convergence rate in terms of number of iterations for the six simulated cases outlined in Table V for three WPs (WP1, WP2, and WP3, respectively). These figures demonstrate the convergence of $\Delta\dot{V}_1$ and $\Delta\dot{V}_2$ at the IBRs terminal within the SS-SC solver. All cases converge in less than five iterations.

To evaluate the solver's ability to handle scenarios where the PF equations become unsolvable due to converter current limitations, we analyzed a test case involving all three WPs operating in GFL mode within the EPRI 120kV full circuit. This case failed to converge, primarily due to the limited reactive power support provided by the GFL converters during fault. When the converter currents are significantly constrained, their reactive power capability becomes inadequate to sustain bus voltage and transmit active power through the lines. As a result, the system enters a state where the PF equations cannot be solved. This reflects the physical phenomenon of grid instability under fault conditions, where insufficient reactive power support from GFL converters leads to voltage collapse and instability. The successful simulation of this scenario demonstrates the proposed method's capability to analyze and capture such instability events.

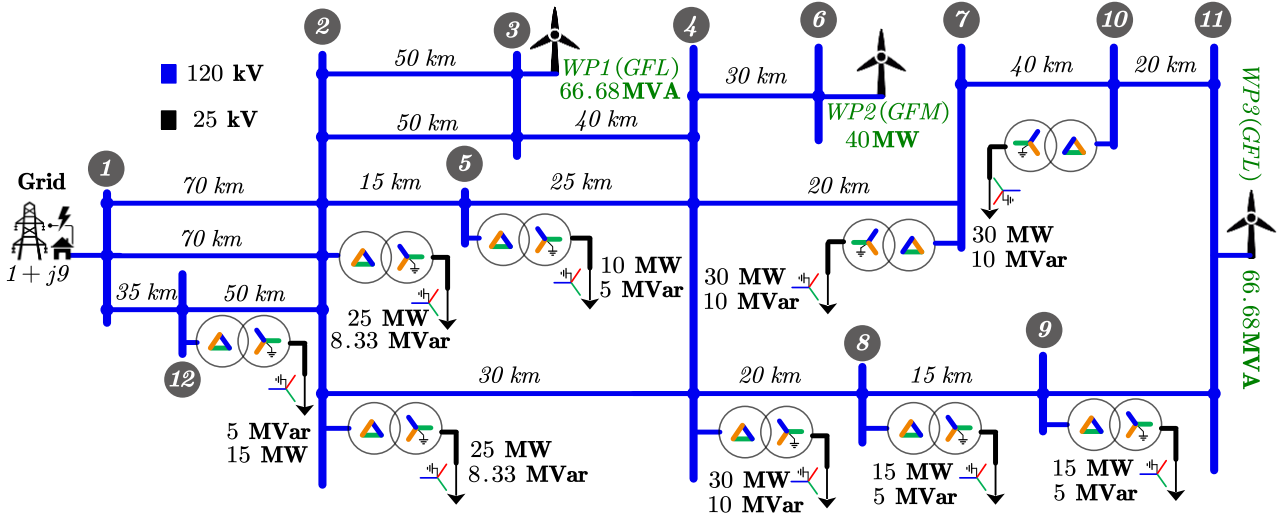

Fig. 9. 120 kV EPRI benchmark (Full Circuit)

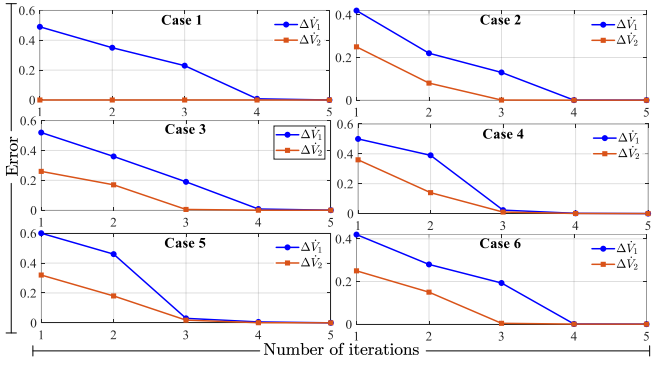

Fig. 10. Number of iterations for $\Delta \dot{V}_1$ and $\Delta \dot{V}_2$ in SS-SC solver (WP1)

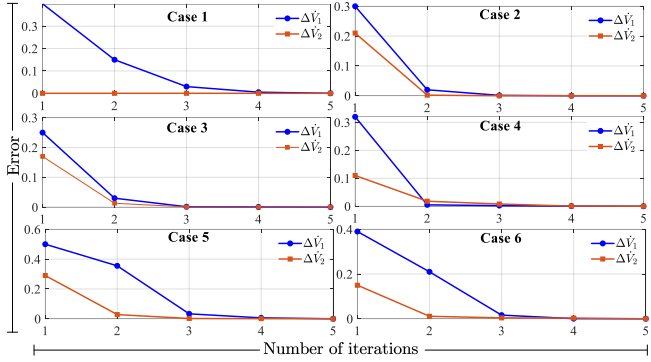

Fig. 11. Number of iterations for $\Delta \dot{V}_1$ and $\Delta \dot{V}_2$ in SS-SC solver (WP2)

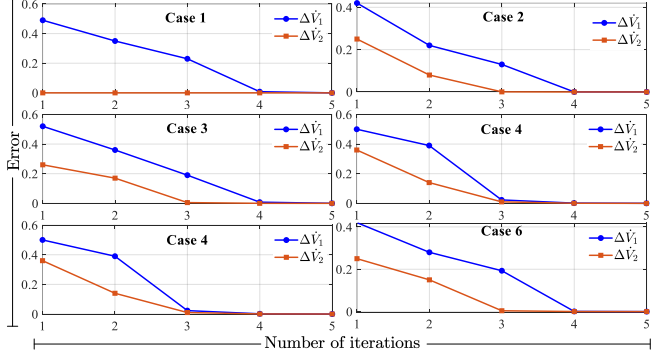

Fig. 12. Number of iterations for $\Delta \dot{V}_1$ and $\Delta \dot{V}_2$ in SS-SC solver (WP3)

## VI. Conclusion

This paper details the development of a multiphase PF solver incorporating sequence domain IBR model, which offers enhanced accuracy compared to the traditional generator-based IBR model. A comparative analysis with time domain EMTP simulation demonstrates the superior performance of the proposed sequence domain IBR-PF model. The classical IBR model shows a maximum voltage magnitude error of 0.176%, whereas the developed IBR-PF model reduces this error to 0.0453%. For current magnitudes, the proposed IBR model achieves error percentages of 0.034%, 0.027%, and 0.001% in the zero sequence, positive sequence, and negative sequence, respectively. In contrast, the classical IBR generator model (voltage source behind impedance) exhibits significantly higher errors of 17.75%, 1.81%, and 31.32% for these sequences.

The developed IBR model was integrated into the SS-SC solver. Various FRT strategies for GFM were reviewed, and CSM and VIC-based FRT approaches were improved through unique sequence current division. Phasor-domain versions were developed, with improved FRT strategies to fully utilize converter capacity while complying with both the German grid code and IEEE-2800 2022 std. Numerical comparisons with EMTP time domain simulations show that the proposed SS-SC solver offers significant computational efficiency, running 1000 times faster while maintaining an error margin below 3%.

The enhanced speed of the solver will facilitate the rapid generation of comprehensive datasets encompassing diverse fault scenarios. These datasets will be utilized to parameterize the fault characteristics of IBRs in future studies. In addition, the proposed solver can be used in transmission planning studies to identify future scenarios having low fault level conditions. This will help to identify system strength issues in different regions of the national transmission system with a high proportion of IBRs.

### Acknowledgement

The authors would like to acknowledge the support provided by the Knowledge Transfer Partnership (KTP) project by Innovate UK, partnership reference number TNEI-13688, in facilitating this research.

## BIOGRAPHIES

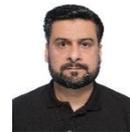

**Dr Zahid Javid** received his BS in Electrical Engineering with honors from University of Engineering and Technology Lahore, Pakistan, in 2016, MS in power system automation from Shandong University, China, in 2019 under CSC Scholarship, and PhD from Hong Kong Polytechnic University, Hong Kong, under prestigious Hong Kong PhD Fellowship Scheme (HKPFS) in August 2023. He worked as a Postdoctoral Researcher at the same institution for 6 months. Currently, he is working as a Power System Innovation Scientist at TNEI & Glasgow Caledonian University as part of a prestigious Knowledge Transfer Partnership Project (KTP) funded by Innovate UK. He is an active contributor to Global Circuit Catalogue IEEE Task Force. His research interests are at the forefront of modern power systems, encompassing power system modelling and analysis, and power flow solutions for modern power grids.

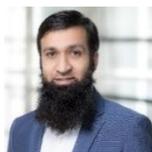

**Dr Firdous UL Nazir** (M'18-SM'22) received the B. Tech. degree in electrical engineering from the National Institute of Technology, Srinagar, India, in 2012, the M. Tech. degree in electrical power systems from Indian Institute of Technology, Roorkee, India, in 2015, and the PhD degree from Imperial College London, U.K. in 2020. He is currently working as a Senior Lecturer in Electrical Power Engineering with the Department of Electrical and Electronic Engineering of Glasgow Caledonian University, Glasgow, U. K. His current research interests include distribution system modelling, operation and control, and optimization theory.

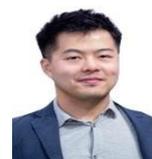

**Dr Wentao Zhu** received his BEng (Hons) and PhD degrees in Electrical and Electronic Engineering from The University of Manchester in 2015 and 2019, respectively. He is an active contributor to the CIGRE D2.56 Working Group, the IEEE Smart Cities Committee, and several IEEE Task Forces. He is a UK Chartered IT Professional (CITP) and an artificial intelligence (AI) data specialist. Currently, he is working as a Senior Power System Engineer at the National Energy System Operator (NESO), UK. With over a decade of experience across consultancy, academia, and system operators, he has developed deep expertise in the design and assessment of whole energy systems and interconnected critical infrastructures, including power, communication, and water networks

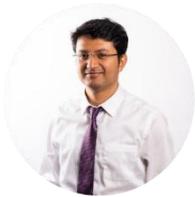

**Dr Diptargha Chakravorty** is a Principal Consultant with Siemens Energy, renowned for his expertise in power system stability and controller interaction studies. Prior to his current role, Dr Chakravorty served as the Head of Innovation at TNEI, where he collaborated closely with UK Network companies on numerous innovation projects, providing technical leadership. His contributions extend to the international arena as an active member of several working groups, including the recent CIGRE B4/C4.97 and the ESIG Large Load Task Force.

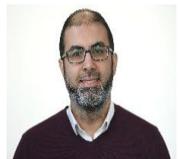

**Dr Ahmed Aboushady** received his BEng (Hons) degree and MSc in Electrical and control Engineering from the Arab Academy for Science and Technology (Alexandria, Egypt) in 2005 and 2008 respectively. Following that, he received his PhD degree from the Department of Electrical and Electronics Engineering at the University of Strathclyde (Glasgow, UK) in 2013, in power electronic systems. Currently, he is a Senior Lecturer (Associate Professor) at Glasgow Caledonian University and the Deputy Director of the Smart Technology Research Centre. Dr. Aboushady led/co-led research projects with total funding awarded of £1.5M from Innovate UK, European Commission, Energy Technology Partnership (ETP) and Net Zero Technology Centre (NZTC). He has over 50 publications in refereed journals/conferences, 1 book, 2 book chapters and 1 patent (PCT/GB2017/051364) and holds Associate Editor role of IEEE Access Journal since 2019. His research interests are in renewable energy systems integration, smart grids, digital twins of power networks and high voltage dc transmission.

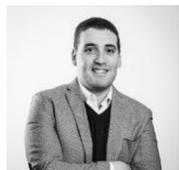

**Dr Mohamed Galeela** received his B.Sc. and M.Sc. degrees in Electrical Power Engineering from Cairo University, Egypt, in 2010 and 2013, respectively, and his Ph.D. from the University of Manchester, U.K., in 2019. He is currently a Principal Power Systems Innovation Consultant within the Energy Networks team at TNEI Services Ltd, Manchester, U.K., where he leads multiple innovation projects focused on the resilience of electric power systems against natural hazards, Black Start services from distributed energy resources (DERs), and grid-forming control for multi-terminal DC networks in offshore wind farms. In addition to his industry role, Dr. Galeela is a visiting researcher at the University of Manchester and an assistant professor (on leave) from Cairo University, Egypt.